\documentclass[traditabstract]{aa} 
\usepackage{graphicx}
\usepackage[varg]{txfonts}
\usepackage{rotating}
\usepackage{natbib}
\usepackage{breqn}

\begin{document}
    \title{The young binary HD$\,$102077: Orbit, spectral type, kinematics, and moving group membership\thanks{Based on observations made with ESO Telescopes at the La Silla Paranal Observatory under programme IDs 084.D-0669(A), 088.C-0753(A), 091.D-0494(A), 53.7-0107, 091.A-9013(A).}}

   \author{ Maria W\"ollert \inst{1} \and
            Wolfgang Brandner\inst{1} \and
 	    Sabine Reffert \inst{2} \and
	    Joshua E. Schlieder\inst{1} \and \\
 	    Maren Mohler-Fischer\inst{1} \and 
 	    Rainer K\"ohler\inst{1} \and 
            Thomas Henning\inst{1}
	   }

   \institute{ Max-Planck-Institut f\"ur Astronomie, K\"onigstuhl 17,
               69117 Heidelberg, Germany\\
               \email{{woellert@mpia.de}} \and
	       Landessternwarte, K\"onigstuhl 12, 69117 Heidelberg, Germany
	      }

   \date{Received ---; accepted ---}

    \abstract{The K-type binary star HD$\,$102077 was proposed as a candidate member of the TW Hydrae Association (TWA) which is a young ($5-15\,$Myr) moving group in close proximity ($\sim 50\,$pc) to the solar system. The aim of this work is to verify this hypothesis by different means. We first combine diffraction-limited observations from the ESO NTT $3.5\,$m telescope in SDSS-i' and SDSS-z' passbands (three epochs) and ESO $3.6\,$m telescope in H-band (one epoch) with literature data to obtain a new, amended orbit fit of the visual binary. We then estimate the spectral types of both components from the i'-z' colours and reanalyse the Hipparcos parallax and proper motion taking the orbital motion into account. Moreover, we use two high-resolution spectra of HD$\,$102077 obtained with the fibre-fed optical echelle spectrograph FEROS at the MPG/ESO $2.2\,$m telescope to determine the radial velocity and the lithium equivalent width of the system. Finally, we use all the information to discuss the kinematic properties of HD$\,$102077 and to estimate the age of the system.

    The orbital elements of the HD$\,$102077 trajectory are well {constrained} and we derive a total system mass of $2.6 \pm 0.8\,$M$_{\sun}$ and a semi-major axis of ${14.9 \pm 1.6}\,$AU. From the i'-z' colours we infer an integrated spectral type of K2V, and individual spectral types of $K0 \pm 1$ and $K5 \pm 1$ for primary and secondary, respectively. The radial velocity corrected {for} the orbital motion of the system is $17.6 \pm 2\,$km/s. Even though the parallax determination from the Hipparcos data is not influenced by the orbital motion, the proper motion changes to $\mu_{\alpha}*\cos{\delta} = -137.84 \pm 1.26$~mas\,yr$^{-1}$ and $\mu_{\delta} = -33.53 \pm 1.45$~mas\,yr$^{-1}$. With {the resultant} space motion, the probability of HD$\,$102077 being a member of TWA is less than ${1\%}$. Furthermore, the lithium equivalent width of $200 \pm 4\,$m\AA $\,$ is consistent with an age between $30\,$Myr and $120\,$Myr and thus older than the predicted age of TWA. The comparison of HD$\,$102077's temperature and luminosity to isochrones supports this result. In conclusion, HD$\,$102077's age, galactic space motion, and position do {not} fit TWA {or} any other young moving group.}


   \maketitle
%

\section{Introduction}

The TW Hya association (TWA) is a loose group of stars {\citep{Gregorio-Hetem92}} that have common {galactic kinematics}, age, and origin \citep{Kastner97, Weinberger13}. The association has $22$ high probability members with individual distances between $28\,$pc and $70\,$pc \citep{Torres08}. The age of its members has been estimated using different indicators of youth
such as activity and lithium abundance as well as their position in the Hertzsprung-Russell diagram and is about $10 \pm 5 \,$Myr \citep{Weinberger13}. Young, nearby associations {like} TWA are of great importance to understanding the local star forming history, and provide a sample of young stars, brown dwarfs, and planets for high-resolution studies. Therefore, several groups have been looking for additional members \citep{Schlieder10, Schlieder12, Zuckerman11, Shkolnik12, Rodriguez13, Malo13, Moor13}.

The binary HD$\,$102077 {(also called HIP 57269, V 838 Cen, and RST 3558AB)} was proposed as a candidate member of TWA based on its kinematics by \cite{Makarov01}. Kinematic candidate members, however, need to be verified by spectroscopic measurements. The first spectroscopic follow up was done by \cite{Song02}, who measured the Li$\,6708\,$\AA $\,$ strength of HD$\,$102077. They find a fairly large equivalent width of $196\,$m\AA, which is a strong indication of a young system. The authors argue, however, that an age of more than $30\,$Myr is more likely and thus do not consider HD$\,$102077 as a new member of TWA. Apart from the lithium absorption line, HD$\,$102077 shows strong X-ray emission which also hints at a young age \citep{Makarov03}. The WISE data \citep{Cutri12} shows, however, no infrared excess and thus no sign of a disk. Even though HD$\,$102077 is a young star, several research groups have assigned it low metallicity values. \cite{Holmberg07} used the Stroemgren UVBY photometry by \cite{Olsen94} to derive a [Fe/H] of $-0.9$. \cite{Casagrande11} redid their analysis and find [Fe/H]$=-0.76$. An independent metallicity measurement was done by \cite{Randich93}. They fitted spectral templates to high-resolution spectra with $R=50000$ over the region of $\approx 6680-6730\,$\AA $\,$  and obtained [Fe/H]$=-0.4$. The photometric determination of the metallicity content is not very accurate because of HD102077’s binarity and variability and this may be the cause of the discrepancy. For instance, \cite{Cutispoto90} measured the $UBVRI$ colours over several photometric phases and observed that $U-B$ varies by $0.04\,$mag and that the star gets bluer at light maximum. Even stronger variations have been observed by \cite{Udalski85}. They find brightness changes of $0.08\,$mag with a probable period of $1.84\,$ or $2.2\,$days. {Since the V-band lightcurves at different epochs differ greatly in shape and mean magnitude \citep{Cutispoto90, Cutispoto93, Cutispoto98} the starspot size and distribution seem to vary considerably. Duplicity of one of the two components could also add to the observed variations.}

\cite{Konig03} measured the radial velocity of the system and find that the space motion is quite different from the average space velocity of TWA. Nevertheless, they point out that the distance of HD$\,$102077, its location in the sky, its spectral type, its position in the H-R diagram, its $v \sin i$, and X-ray emission are very similar to other confirmed TWA members. {This group also presents evidence of a third component in the system. The proposed star is at a distance of $8.46''$ from the primary and it shows calcium and lithium absorption. The lithium EW of $0.18 \pm 0.2\,$\AA $\,$ and a spectral type of $K5 \pm 1$ point to a similar age to the A component. We checked the imaging archives to see whether there is an indication of common proper motion. The stars do seem to move together with respect to the background stars in the field when comparing an R-band image from $1983$ to an R-band image from $1992$ and the 2MASS K-band image from $1999$. The psfs are, however, blended in all these images and thus no quantitative proper motion measurement is possible. Nevertheless, the calcium and lithium absorption features as well as the likely common proper motion is strong evidences of a bound system. }

{The integrated spectral type of HD$\,$102077 of $K0/1Vp$ was first determined by \cite{Anders91} by high-dispersion observations of the Li $6707$ \AA \, doublet and \cite{Konig03} find $K1/2$ using DFOSC at the $1.54\,$m Danish telescope.} The spectral type of the {primary} components is, however, less clear. \cite{Fabricius00} analysed the Hipparcos Tycho data and find that the brighter component has a redder B-V colour, corresponding to spectral types K4 and K2, but they do not give an explanation for this unexpected result. {\cite{Cutispoto98} uses the UBVRI colours of the integrated system and $\Delta H_p$ from the Hipparcos satellite catalog to classify HD$\,$102077 as consisting of a $K0/1V$ and a $K5V$.}

The source HD$\,$102077 is a visual binary star and thus allows the determination of the total system mass. The first orbital parameters were derived by \cite{Heintz86}. His best fit to data from seven epochs is a low eccentricity orbit with a period of $32.35\,$years and a semi-major axis of $0.28''$. At that time, the parallax of HD$\,$102077 was not known. Therefore, Heintz used Kepler's Third Law and the mass-luminosity relation to estimate the dynamical parallax. He obtained $42.3\,$pc and deduced a total system mass of $1.6\,M_{\sun}$. 

In this paper, we present the results of our astrometric, photometric, and spectroscopic study of HD$\,$102077. We derive a new orbit fit of the binary system, and we determine the i'-z' colours and from these the spectral types of the individual components. We measure the radial velocity and the lithium equivalent width of the system. Finally, we use these results to discuss a possible TWA membership of HD$\,$102077. 

\section{Observations and data reduction}

\begin{table}[htb]
    \caption[]{Astrometric measurements of HD$\,$102077} 
    \label{data}
      \begin{center}
	\begin{tabular}{l l l l}
	  \hline
	  \hline
	Date      &  PA [$^{\circ}$] &     $\rho\,$['']           &  Ref \\
	  \hline
	1929.065  &  $348.9  \pm 5.0$  &  $0.24  \pm 0.1$    &  Ro  \\
	1949.529  &  $270.0 \pm 5.0$  &  $0.17  \pm 0.1$    &  Ro  \\
	1964.32   &  $60.5 \pm 5.0$  &  $0.19  \pm 0.1$    &  Ho  \\
	1976.125  &  $164.4  \pm 5.0$  &  $0.14  \pm 0.1$    &  Wo  \\
	1980.231  &  $243.5  \pm 5.0$  &  $0.14  \pm 0.1$    &  Wo  \\
	1983.168  &  $264.6  \pm 5.0$  &  $0.19  \pm 0.1$    &  Wo  \\ 
	1985.34   &  $278.5  \pm 5.0$  &  $0.33  \pm 0.1$    &  He  \\
	1989.3059 &  $298.7  \pm 0.2$  &  $0.416   \pm 0.003$  &  Mc  \\
	1990.3462 &  $302.9  \pm 0.2$  &  $0.42  \pm 0.003$  &  Ha  \\
	1991.25   &  $306 \pm 1.0$  &  $0.43  \pm 0.006$  &  Hi  \\
	1994.504  &  $320.2  \pm 0.7$  &  $0.418   \pm 0.004$  &  *   \\
	2001.0825 &  $352.9^a  \pm 0.7$  &  $0.335   \pm 0.004$  &  Ma  \\
	2009.2598 &  $115.5  \pm 0.3$  &  $0.1795 \pm 0.001$  &  To  \\
	2010.08   &  $135.9  \pm 0.5$  &  $0.183   \pm 0.003$  &  *   \\
	{2011.0373} &  ${157.3^a} { \pm 0.2}$  &  ${0.1874} {  \pm 0.002}$  &  {Hk}  \\
	2012.011  &  $177.3  \pm 0.5$  &  $0.196   \pm 0.003$  &  *   \\
        2013.3042 &  $202.9  \pm 0.7$  &  $0.21  \pm 0.003$  &  *   \\	
	  \hline
	\end{tabular} 
      \end{center}
    \begin{quote}
	    Ro: \cite{Rossiter55}, Ho: \cite{Holden65}, Wo: \cite{Worley78}, He: \cite{Heintz86}, Mc: \cite{McAlister90}, Ha: \cite{Hartkopf93}, Hi: \cite{HipCat97}, Ma: \cite{Mason11}, To: \cite{Tokovinin10}, Hk: \cite{Hartkopf12}, *: This work, a: position angle corrected by $180^{\circ}$
    \end{quote}
\end{table}

\subsection{Literature data}

The source HD$\,$102077 has been observed in the optical with various telescopes and instruments since 1929 (see Table~\ref{data}). All measurements until $1985$ were done with a filar micrometer. Because of the small separation of the two components close to the resolution limit of the telescopes used, we adopt a high uncertainty in the separation of $0.1''$ and $5^{\circ}$ in the position angle for all micrometer measurements. The uncertainties of the later measurements are the values reported by the authors in the papers or in follow-up studies. We note that the binary was not resolved by Worley in 1972 nor by Heintz in 1984 \citep{Heintz86}.

\subsection{AO observation with COME-on+/SHARPII}

The observations of HD$\,$102077 in the H-band were obtained on July 2, 1994, with the adaptive optics (AO) system ComeOn+ \citep{Rousset94} and the near-infrared camera Sharp2 \citep{Hofmann92} at the ESO $3.6\,$m telescope in La Silla, Chile. For sky subtraction, HD$\,$102077 was observed in two different quadrants of the detector with $30$ individual frames each. Exposure times of the single frames were $1\,$s. The star positions for both configurations were measured with IRAF imexamine assuming a two-pixel object radius. Their values were averaged, and the standard deviation was used to estimate measurement uncertainties. 

Astrometric calibrations were achieved through observations of the astrometric binary IDS 17430S6022 \citep{vanDessel93}. These observations yield an image scale of 50.20 $\pm $ 0.12 mas/pixel and show that the y-axis of the detector was aligned with the north direction to within $\pm $0.20 deg.

\subsection{Lucky Imaging with AstraLux}

Three additional epochs were obtained using the Lucky Imaging camera AstraLux Sur at the ESO $3.5\,$m New Technology Telescope at La Silla \citep[see][for more information on AstraLux]{Hormuth08, Hippler09}. Lucky Imaging is a passive technique which enables nearly diffraction-limited observations on 2- to 4-meter telescopes. The seeing is reduced by taking a series of many short exposures of the target and combining only the least distorted frames to the final image. 

\begin{figure}[htb]
    \begin{center}
	\includegraphics[width=7.0cm,angle=0]{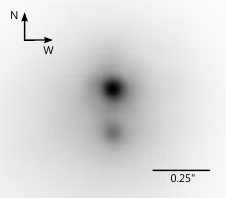}
    \end{center}
    \caption[z' filter image of 2012]{z' filter image of HD$\,$102077 with AstraLux at La Silla observatory from January $5$, 2012. The image scale is linear.} 
    \label{z_image}
\end{figure}

We observed HD$\,$102077 on February $7$, 2010; on January $2$, $4$, and $5$, 2012; and on April $21$, 2013. At each time we imaged HD$\,$102077 in i'- and z'-passband except on January $4$, 2012, when we only took a z'-band image, and January $2$, 2012 when we observed HD$\,$102077 in z'-band twice. The z'-band image from January $5$, 2012, is shown in Fig.~\ref{z_image} {as an example}. The exposure time of individual frames was $30\,$ms in 2010 and $15\,$ms on the other dates. In order to match a total integration time of $300\,$s, the number of integrations was set to $10000$ and $20000$, respectively. 

The single frames were corrected for electronic offset and differences in pixel sensitivity using a masterflat and -bias, respectively. After noise filtering and resampling the pixel scale from $30.5\,$mas/pixel to $15.25\,$mas/pixel, the Strehl ratio of the point spread function was measured. Only $1\%$ of the images with the highest Strehl ratio were then combined to the final image. The assembling is based on the Drizzle algorithm \citep{Fruchter02}. It shifts the selected images such that the brightest pixel of the reference star is always positioned at the same pixel coordinates.

Astrometric reference stars in the open cluster NGC$\,3603$ were observed each night to calibrate the detector rotation and pixel scale. We measured the detector position for five stars of NGC$\,3603$ with IRAF imexamine, calculated the {separation} and rotation angle pairwise, and compared these values to data from high-quality astrometric HST/WFP$2$ observations \citep{Rochau10}. This analysis yielded a plate scale of $15.25 \pm 0.03$ mas/px, $15.25 \pm 0.01$ mas/px, and $15.24 \pm 0.02$ mas/px and a detector rotation of $1.8^{\circ} \pm 0.2^{\circ}$, $2.1^{\circ} \pm 0.2^{\circ}$, and $3.8^{\circ} \pm 0.2^{\circ}$ west of north in 2010, 2012, and 2013 respectively. Additionally, twilight sky-flats and bias were measured in both filters at the beginning of each observing night. 

The positions of the binary components as well as their magnitude differences in the i' and z' filters were obtained by fitting a model psf to the data. The co-adding of images of short exposures with random atmospheric distortions is described by \citep{Hormuth08}:
\begin{dmath}
  \textrm{PSF}_{\textrm{\tiny{obs}}}(r) = W \left( \frac{1}{r^2/\sigma^2_m +1} \right)^{\beta} \\ + (1-W) \left( \textrm{PSF}_{\textrm{\tiny{th}}}(r) \ast \frac{1}{\sqrt{2 \pi \sigma^2_g}} \exp \left( -\frac{r^2}{2\sigma^2_g} \right) \right).
\end{dmath}
On the right-hand side $\textrm{PSF}_{\textrm{\tiny{th}}}$ is the ideal point spread function without atmospheric turbulences and telescope aberrations, which is then convolved with a Gaussian blurring function of parameter $\sigma_g$. To obtain the observed PSF, a Moffat profile with parameters $\beta$ and $\sigma_m$ is added to the blurred theoretical profile; $W$ weights the two PSF components. Our code determines all parameters simultaneously by a Levenberg-Marquardt least-squares fit to both binary components. 

In order to unveil systematic errors of the code we constructed binaries from seven single stars that we imaged on the same night. The synthetic binaries had similar separation and brightness ratios to HD$\,$102077, and were introduced at position angles of $0^{\circ}, 45^{\circ}, 90^{\circ}, 135^{\circ}, 180^{\circ},$ and $270^{\circ}$. We compared the fit results with the set system properties and found no systematic offset in separation, position angle, and flux ratio.

In addition to the flux ratio of the binary component aperture photometry with IRAF apphot was used for our observations from 2010 to determine the i' and z' magnitudes of the integrated system. We chose an aperture radius of $21\,$px centred on the primary star to ensure that most of the flux of both components is covered. The sky background was measured in an annulus from $25\,$px to $35\,$px. The corresponding photon flux is then calculated taking into account the system gain, the exposure time, the mirror area, and the quantum efficiency of our CCD in i' and z' passband which is $0.8$ and $0.4$, respectively. For an absolute calibration of the instrument we observed two astrometric standard stars at a comparable airmass during the same night and analysed them in the same way.  
To calibrate our instrument we only considered offsets between our instrumental magnitude and the standard value given in literature and no colour terms.

\subsection{High-resolution spectroscopy with FEROS}

Two high-resolution spectra of HD$\,$102077 were obtained on July 26 and July 27, 2013, with the fibre-fed optical echelle spectrograph FEROS \citep{Kaufer98} at the MPG/ESO $2.2\,$m telescope at La Silla Observatory in Chile. FEROS covers the whole optical spectral range from $3600\,$\AA $\,$ to $9200\,$\AA $\,$ and provides a spectral resolution of $\approx 48000$. The fibre aperture of FEROS is $2\,$arcsec and thus contains the light of both components. The exposure time of each spectrum was $1500\,$s and resulted in a signal-to-noise ratio of $223$ and $192$ at $5450\,$\AA, respectively. The spectra were taken in the object-cal mode and were reduced with the online Data Reduction System (DRS) at the telescope. The reduction includes detection of spectral orders, wavelength calibration and background subtraction, flatfield correction and order extraction, as well as merging the individual orders after rebinning the reduced spectra to constant wavelength steps. Since the FEROS pipeline applies the barycentric correction inaccurately \citep{Mueller13}, we recalculated the barycentric correction with the IDL routine baryvel.pro including the rotation of Earth at the observatory.

The stellar parameters of HD$\,$102077 are determined by fitting synthetic spectra to the stellar spectrum using the spectral synthesis code Spectrocopy Made Easy, \textit{SME} \citep{Valenti96}. We use these stellar parameters to compute a template spectrum which we then cross-correlate with the measured spectrum to determine the radial velocity of the system. Furthermore, we measure the Li$\,6708\,$\AA $\,$ equivalent width.

\section{Physical properties of the HD$\,$102077 binary}

\subsection{Orbit determination}

The orbit of a binary system can be characterized by the relative motion of the secondary around the primary. Even though we only see a projection of the movement on the celestial sky, a sequence of observations at many different epochs, preferably covering more than one complete orbit, makes it possible to determine all orbital elements. The astrometric data of HD$\,$102077 covers more than two full orbits (see Table~\ref{data}). 

To find the best fit to that data we employ a grid-search in eccentricity $e$, period $P$, and time of periastron $T_0$. For every set of these values we solve the Kepler equation and we get the eccentric anomaly $E$ for every point in time $t$ that has been measured: 
\begin{eqnarray*}
E - e \sin{E} = \frac{2 \pi}{P}(t-T_0).
\end{eqnarray*}
Then the elliptical rectangular coordinates of the real orbit $X$ and $Y$ are calculated at every epoch,
\begin{eqnarray*}
X &=& \cos{E}-e \\
Y &=& \sqrt{1-e^2} \cdot \sin{E},
\end{eqnarray*}
and $X$ and $Y$ are connected with the measured projected points $x$, $y$ via the four Thiele-Innes coefficients $A$, $B$, $F$, $G$:
\begin{eqnarray*}
x &=& A \cdot X + F \cdot Y \\
y &=& B \cdot X + G \cdot Y. 
\end{eqnarray*}
The Thiele-Innes coefficients in turn depend on the orbital elements, the semi-major axis $a$, the angle between node and periastron $\omega$, the position angle of the line of nodes $\Omega$, and the inclination $i$:
\begin{eqnarray*}
A &=& a \, (\cos{\omega}\cos{\Omega}-\sin{\omega}\sin{\Omega}\cos{i}) \\
B &=& a \, (\cos{\omega}\sin{\Omega}+\sin{\omega}\cos{\Omega}\cos{i}) \\
F &=& a \, (-\sin{\omega}\cos{\Omega}-\cos{\omega}\sin{\Omega}\cos{i}) \\
G &=& a \, (-\sin{\omega}\sin{\Omega}+\cos{\omega}\cos{\Omega}\cos{i}).
\end{eqnarray*} 

\begin{figure}[htb]
    \begin{center}
	\includegraphics[width=8.5cm,angle=0]{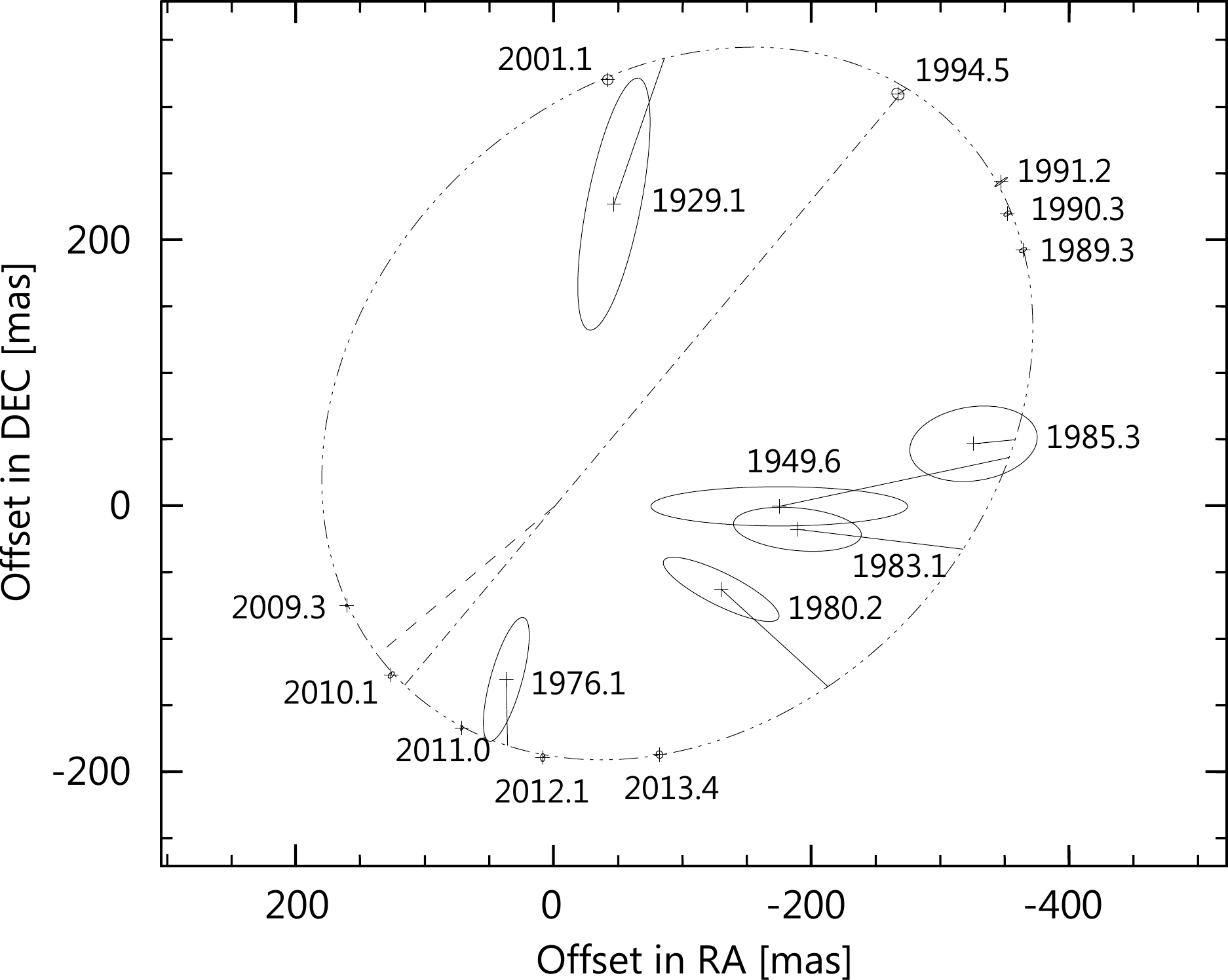}
    \end{center}
    \caption[orbit]{Best fitting orbit of HD$\,$102077 in the rest frame of the primary. The observed positions are marked by their error ellipses and lines connecting the observed and calculated positions at the time of the observation. The dash-dotted line indicates the line of nodes, the dashed line the periastron. We note that the micrometer measurements all tend to estimate a smaller projected separation.} 
    \label{orbit_fit}
\end{figure}

Using all epochs $A$, $B$, $F$, and $G$ are calculated using {singular value decomposition}. Finally, the orbital elements can be computed. The goodness of every fit is measured with the $\chi^2$ test. The minimal $\chi^2$ corresponds to the best fit. The best orbital solution for HD$\,$102077 has a reduced $\chi^2$ of $2.6$ and can be seen in Fig.~\ref{orbit_fit}; the corresponding orbital elements can be found in Table~\ref{orbit}. Uncertainties were estimated by analyzing the $\chi^2$ function around its minimum. The errors given in Table 2 represent the $68\%$ confidence interval (see \cite{Kohler08} for more details). 

The semi-major axis of ${14.9 \pm 1.6} \,$AU and the system mass of $2.6 \pm 0.8 \,$M$_{\sun}$ were calculated using the revised Hipparcos parallax of $20.6 \pm 2.1\,$mas. The uncertainties of these values are clearly dominated by the distance uncertainty. We note that we excluded the data point from $1964$ for our orbital fit. It shows a significant and unexplainable offset compared to all other data points and we have no means to recheck its derivation. {We also added $180^{\circ}$ to the position angle measurements from $2001$ and $2011$ in order to match the other observations.}

\begin{table}
    \caption{{Parameters of the best orbital solution.}}
    \label{orbit}
    \renewcommand{\arraystretch}{1.3}
    \begin{center}
	\begin{tabular}{lr@{}l}
	      \noalign{\vskip1pt\hrule\vskip1pt}
	      \noalign{\vskip1pt\hrule\vskip1pt}
	      Orbital Element				      & \multicolumn{2}{c}{Value} 		\\
	      \noalign{\vskip1pt\hrule\vskip1pt}
	      Date of periastron $T_0$	(JD)		      & $2455142$ & $\,^{+   4}_{   -6}$	\\
	      Period $P$ (years)                              & $   35.5$ & $\,^{+0.3}_{-0.3}$		\\
	      Semi-major axis $a$ (mas)                       & $  306.2$ & $\,^{+1.1}_{-1.4}$		\\
	      Inclination $i$ ($^\circ$)                      & $   28.5$ & $\,^{+0.4}_{-1.3}$		\\
	      Eccentricity $e$                                & $  0.409$ & $\,^{+0.004}_{-0.002}$	\\
	      Argument of periastron $\omega$ ($^\circ$)      & $ 348.60$ & $\,^{+0.09}_{-0.91}$	\\
	      P.A. of ascending node $\Omega$ ($^\circ$)      & $  140.0$ & $\,^{+0.1}_{-1.3}$		\\
	      System mass $M_S$ (M$_{\odot}$/kpc$^3$)	      & $  22834$ & $\,^{+ 265}_{-359}$	        \\			
	      \noalign{\vskip1pt\hrule\vskip1pt}
	      Semi-major axis $a$ (AU)                        & $     14.9$ & $\,^{+   1.6}_{  -1.6}$  	\\
	      System mass $M_S$ (M$_\odot$)                   & $    2.6$ & $\,^{+0.8}_{-0.8}$		\\
	      \noalign{\vskip1pt\hrule\vskip1pt}

	\end{tabular}
    \end{center}
\end{table}

\subsection{Stellar parameters}

\subsubsection{i'-z' colours}

The spectral type of the integrated system as well as the spectral types of the individual components are determined from i'-z' colours which we compare to the spectral type models for main-sequence stars by \cite{Kraus07}. Even though HD$\,$102077 may still be in its pre-main-sequence phase, it is reasonably close to the zero-age main sequence to justify the adoption of main-sequence colours and magnitudes. We derive an integrated spectral type of K2V. To determine the i'-z' colours of the individual components we need to disentangle their signal. This is achieved by fitting a theoretical PSF to both stars simultaneously. Averaging the results of all five z'-band images and three i'-band images we get a flux ratio of $0.377 \pm 0.02$ and $0.325 \pm 0.01$, respectively. Together with the apparent integrated i'- and z'-magnitudes, this yields a spectral type of $K0 \pm 1$ for the primary and $K5 \pm 1$ for the secondary.

\subsubsection{Spectra}

In fitting model spectra to our binary system we find $T_{\textrm{\tiny{eff}}} \approx 5265\,$K, $\log g \approx 4$, Fe/H $\approx -0.37$, and $v \sin i$ $\approx 8.3 \,$km/s. However, these results should be taken with care since we only fit one component to the unresolved binary system. The lines are expected to be broadened from the relative motion of the two components which can be calculated from our orbit fit to be $5 \pm 0.5\,$km/s in July 2013 when the spectra were taken. We find a radial velocity of $19.8\,$km/s with a small uncertainty of $0.1\,$km/s introduced by the measurement and larger uncertainty introduced by the orbital motion. The measured velocity lies between the radial velocity of the primary and the secondary and it depends on the mass ratio and on the flux ratio of the two components. Since the exact values are not known we adopt an uncertainty of $2.5\,$km/s, half of the relative speed, corresponding to the highest possible uncertainty.

We also calculated the radial velocity of the individual components with an IDL analysis package that is based on the 2D cross correlation technique described in \cite{Mazeh02} (C. Bender, private communication). The cross correlation of the two spectra of HD$\,$102077 with one K0 template gives $19.8 \pm 0.4\,$km/s and confirms the above mentioned value. The simultaneous cross correlation with an K0 and an K5 template increases the correlation coefficient significantly and gives an individual radial velocity of $26.5 \pm 0.9\,$km/s for the primary and $12.7 \pm 0.7\,$km/s for the secondary in four different wavelength ranges and in both spectra. The radial velocity difference of about $14\,$km/s of the two components does not agree with the predicted value of $5\,$km/s from our orbit fit. This could be explained either by a third component in the system or by a parallax measurement that is too large by about a factor of $2.8$. The second would, however, lead to an unreasonably high system mass of about $57\,$M$_\odot$. In addition, our reanalysis of the Hipparcos parallax does not suggest an incorrect parallax determination (see section 4.3). 

\subsection{V-band variability}

\begin{figure}[htb]
  \begin{center}
	\includegraphics[width=9.0cm,angle=0]{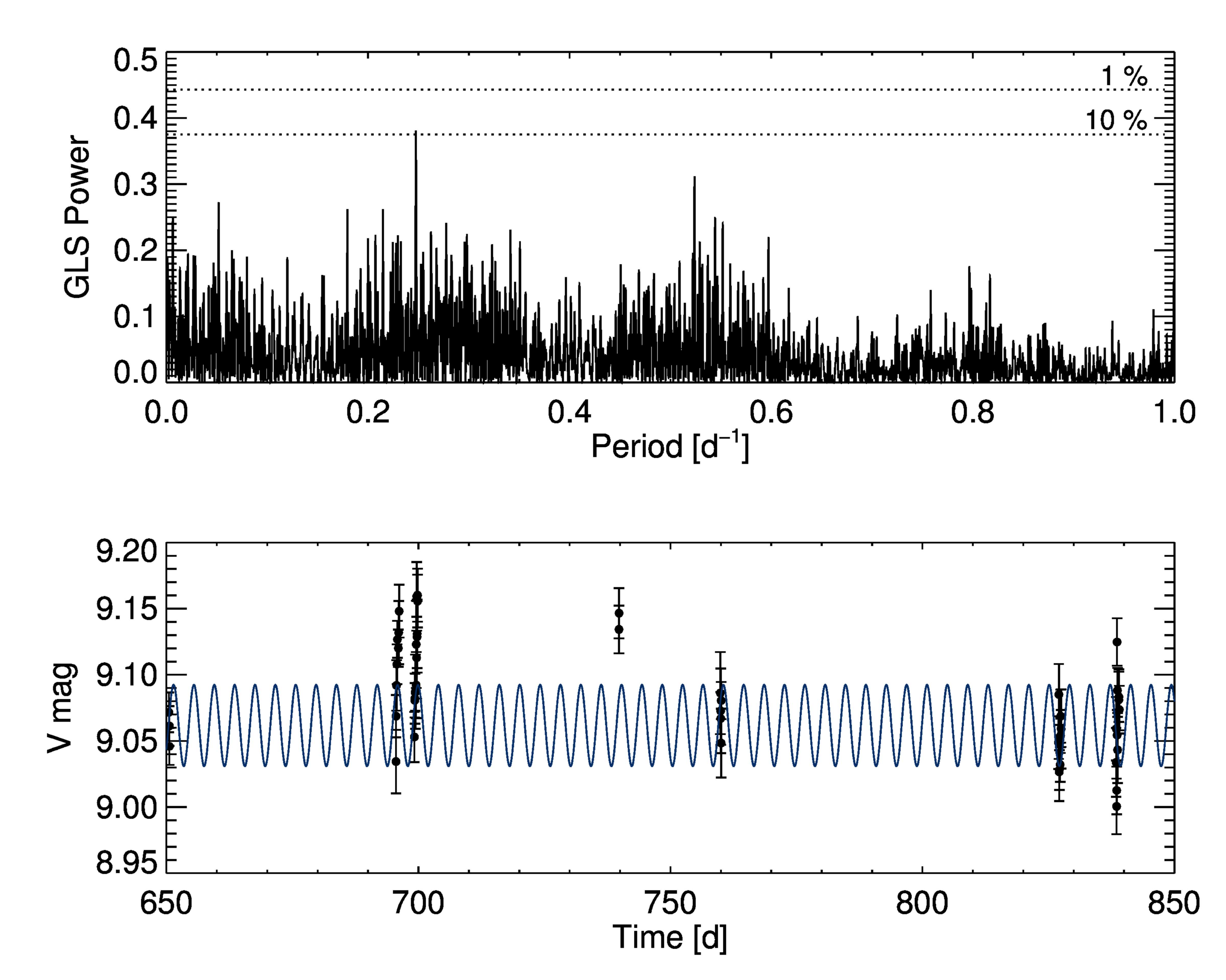}
  \end{center}
  \caption[Generalized Lomb-Scargle periodogram of HD$\,$102077]{Top: Generalized Lomb-Scargle periodogram for Hipparcos V-band data. Hte highest power is reached for a period of $4.04$ days. The false alarm probability of $1 \, \%$ and $10 \, \%$ are indicated by the dotted lines. Bottom: A part of the Hipparcos light curve is overplotted with the most significant periodic signal.} 
  \label{periodogram}
\end{figure}

We analysed the variability of HD$\,$102077 by studying the Hipparcos V-band data. During the mission lifetime the satellite scanned $166$ times over HD$\,$102077. The observations span more than three years and thus have a much longer baseline than all previous measurements. To find repeating signals we calculated the generalized Lomb-Scargle periodogram ~\citep{Zechmeister09} shown in Fig.~\ref{periodogram}. The highest power is obtained by a period of $4.04\,$days followed by a period of $1.8\,$days. These powers are, however, small and are most likely not associated with a real signal. To test whether the maximum peaks arose just by chance we deleted $20\%$ of the data randomly and recalculated the periodogram; this procedure was repeated several times. Most often we find different prominent peaks. In addition, we analysed the light curve with two algorithms which do not assume a sinusoidal signal: a) box-fitting least squares \citep{Kovacs02} which is optimized to find periodic transits, and b) Plavchan \citep{Plavchan08} which makes no assumption about the underlying periodic signal. 
Neither of these algorithms shows a solution with a high power and the most likely periods differ when deleting $20\%$ of data randomly. {This suggests that apart from the brightness variation of HD$\,$102077 due to stellar activity as has also been seen in the shorter period observations \citep{Cutispoto90, Cutispoto93, Cutispoto98} no other signal can be detected in the Hipparcos data.}

\subsection{Age estimate}

\subsubsection{Using isochrones}
\begin{figure}[htb]
    \begin{center}
	\includegraphics[width=9cm,angle=0]{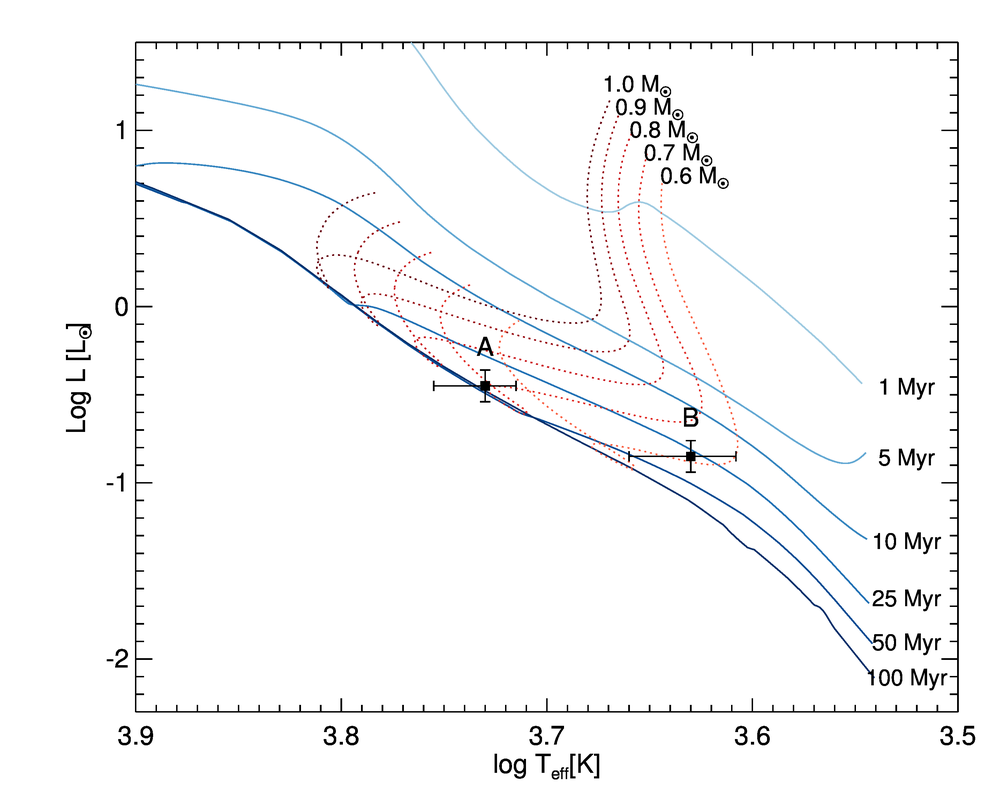}
    \end{center}
    \caption[H-R diagram]{The position of both components of HD$\,$102077 in an H-R diagram along with the Pisa pre-main-sequence tracks and isochrones \citep{Tognelli11} for a metal abundance of $\textrm{Z}=0.005$, a helium fraction of $\textrm{Y}=0.273$, and a mixing length of $\textrm{ML}=1.68$.} 
    \label{HR}
\end{figure}

We use Pisa (FRANEC) pre-main-sequence isochrones \citep{Tognelli11} to estimate the age of HD$\,$102077. The isochrones are computed for a fine grid of mass, age, metallicity, and helium abundance and give the bolometric luminosity as a function of the effective temperature of stars at a certain age. We deduce the effective temperature of HD$\,$102077's two components from our spectral type estimates using the stellar SEDs of \cite{Kraus07}. We find $\log \textrm{T}_{\textrm{\tiny{eff}}}=3.73_{-0.015}^{+0.025}$ for the primary and $\log \textrm{T}_{\textrm{\tiny{eff}}}=3.63_{-0.022}^{+0.030}$ for the secondary. We calculate the bolometric luminosity from the Tycho $V_T$ and $B_T$ colours of $9.44\,$mag and $10.50\,$mag for the primary and $10.56\,$mag and $11.47\,$mag for the secondary. We assume that the uncertainty of the Tycho-magnitudes is dominated by the stellar variability whose $1 \sigma$value is $0.04\,$mag. Using the theoretical calibration from Tycho to Johnson magnitudes \citep{Bessell00} and the bolometric correction from \cite{Voigt88} we find a bolometric magnitude of $5.60_{-0.23}^{+0.21}\,$mag for the primary and $6.33_{-0.23}^{+0.21}\,$mag for the secondary. We now place both components in a H-R diagram (see Fig.~\ref{HR}) together with Pisa pre-main-sequence isochrones. As input parameters we use a metallicity value of $-0.4\,$dex as was determined by \cite{Randich93} and our own FEROS measurement and we assume solar helium mass fraction and mixing length. We overlayed stellar tracks for comparison. Both components lie above the main sequence, with an age between $25 \,$Myr and $100 \,$Myr for the primary, and $15\,$Myr and $50\,$Myr for the secondary.

\subsubsection{Using the lithium\,I $\lambda$6708\,\AA \, equivalent width}
To determine the stellar age independently from the stellar parameters, the equivalent width (EW) of the Li\,I $\lambda6708$\,\AA \, line can be used \citep{Wallerstein69}. The reaction rates of the lithium burning depends on the age of the star and on its mass, and the corresponding effective temperature and young stars exhibit larger Li features than older stars. Although this age estimation method is under discussion because non-steady accretion can heavily affect the Li\,I depletion \citep{Baraffe10}, it is still a commonly used age indicator \citep[e.g.][]{Thomann13}. \\
Since the Li\,I line at $6708$\,\AA \, is blended with an Fe I line, the line was fitted with a Voigt profile to account for the blending. Using the Voigt fit, the EW was derived. By using the SNR of the spectra as a weight, the average Li EW with error was calculated from the individual EWs derived from each single spectrum. Since the binary does not show WISE IR-excess, disk accretion is unlikely to occur and we do not expect veiling \citep{Lynden74, Kenyon87, Bertout88}. The resulting Li\,I EW is $200 \pm 4$\,m\AA \, which agrees well with the previously measured values {for HD$\,$102077} \citep{Song02} {and its probable third, wider component discovered by \cite{Konig03}.}  \\

\begin{figure}
\centering
\includegraphics[width=0.5\textwidth]{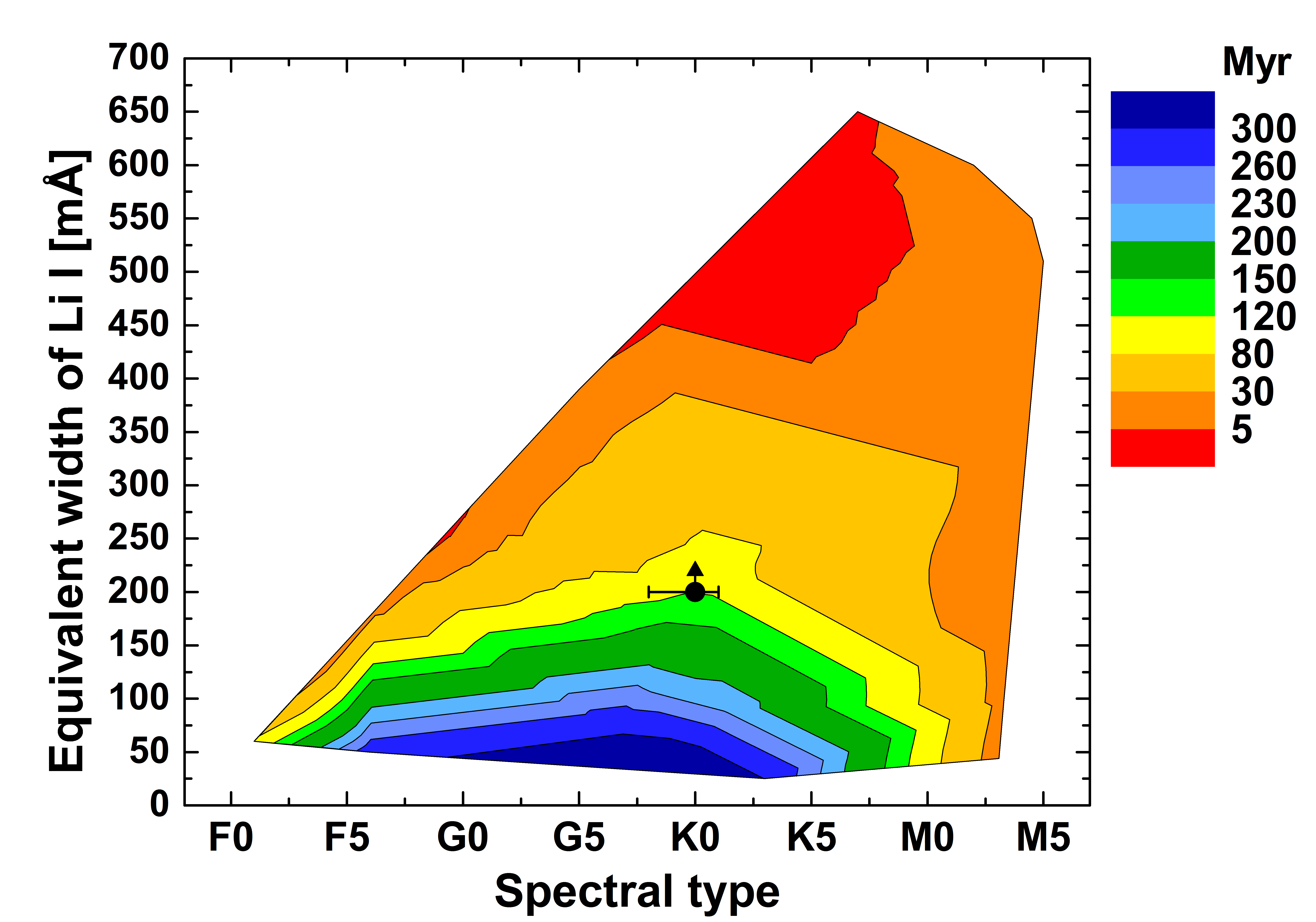}
\caption{Li\,I EW over spectral type. The coloured contours show the age calibration of \citet{Zuckerman04}. The black circle indicates the upper limit of the age of the primary component with spectral type error bars from photometry.}
\label{fig:Li_HD_102077}
\end{figure}

In order to determine the stellar age from the derived Li EW, an age calibration is needed. \citet{Zuckerman04} used eight different stellar associations with calculated Li EWs and independently derived mean age. By doing so, they derived an age calibration of Li\,I EW over (B-V) and spectral type, allowing a simple placement of object measurements in the calibration.\\
Figure~\ref{fig:Li_HD_102077} shows the Li EW measurement of HD$\,$102077 of $200 \pm 4 \,$m\AA\, in the diagram. This corresponds to a maximum age of $120\,$Myr. The minimum age is not as well {constrained} as we measure the EW of the combined signal, and the later K-type secondary star is expected to have a smaller EW than the primary. {The lithium analysis would benefit from taking the binarity of the system into account. Our spectral resolution and the signal-to-noise ratio are, however, not high enough to disentangle the signal of both stars.}

\begin{table}
    \caption{{Summary of photometric and spectroscopic results.}}
    \label{stellar}
    \begin{center}
	\begin{tabular}{l|l}
	      \noalign{\vskip1pt\hrule\vskip1pt}
	      \noalign{\vskip1pt\hrule\vskip1pt}
	      Property					      & Value \\
	      \noalign{\vskip1pt\hrule\vskip1pt}
	      Spectral type primary			      & $K0 \pm 1$ \\ 
	      Spectral type secondary			      & $K5 \pm 1$ \\ 
	      \noalign{\vskip1pt\hrule\vskip1pt}
	      Photometric period			      & none  significant \\
	      \noalign{\vskip1pt\hrule\vskip1pt}
	      Isochrone age (Myr)			      & $> 25$ \\
	      Lithium age (Myr)				      & $< 120$ \\
	      \noalign{\vskip1pt\hrule\vskip1pt}
	      Mean radial velocity (km/s)		      & $17.6 \pm 2$\\
	      $\mu_{\alpha}*\cos{\delta}$ (mas/yr)     	      & $-137.84 \pm 1.26$ \\
	      $\mu_{\delta}$ (mas/yr)                         & $-33.53 \pm 1.45$ \\

	      \noalign{\vskip1pt\hrule\vskip1pt}
	  \end{tabular}
    \end{center}
\end{table}

\section{Discussion}

\subsection{{Additional companions}}

In addition to the wide companion candidate described by \cite{Konig03}, our mass estimate of $2.6 \pm 0.8\,$M$_{\sun}$ may indicate that HD$\,$102077 has another, so far unknown companion; this is considered because the two K-type stars should be less massive than $1.6\,$M$_{\sun}$. This would be in accordance with the classification as RS CVn by \cite{Weiler79}, \cite{Gurzadian92}, and \cite{Pallavicini92}. In addition, the 2D cross correlation shows a large radial velocity difference which is in disagreement with the orbital fit, but could be caused by a third component. In contrast, the V-band variability seems to be linked with stellar activity and HD$\,$102077 has also been classified as BY Draconis by several groups \citep{Udalski85, Alekseev96}. Whether HD$\,$102077 is only active or a spectroscopic binary is {thus} not clear. Further evidence is needed to {answer} this question.

\subsection{Spectral type derivation}

From the i'-z' colours we determined an integrated spectral type of K2V. The spectral type determined with spectroscopic means was $K0/1Vp$ \citep{Anders91} and is thus one subclass earlier. This deviation is common in a spectral classification based on two colours. We found the individual spectral types to be $K0 \pm 1$ and $K5 \pm 1$ {which are in agreement with $K0/1V$ and $K5V$ derived by \cite{Cutispoto98}}. This classification seems to be more reliable than the K4 and K2 determined by \cite{Fabricius00} since the brighter primary star is expected to have the earlier spectral type unless the companion is a white dwarf. We note that the errors of the two spectral types are correlated. An earlier spectral type of the primary star {implies} a later type for the secondary.

\subsection{Revisiting the Hipparcos data for HD$\,$102077}

In the original version of the Hipparcos
Catalog (ESA 1997) the dataset of HD$\,$102077 = HIP$\,$57269 was solved as a two-component system, with identical
proper motions and parallaxes. The separation and position angle are
given in Table$\,$1 and have been included in the visual orbital fit; 
the absolute
astrometry given for HIP$\,$57269 in the main Hipparcos Catalog corresponds
to component A. The solution quality is grade 'A', indicating the highest
possible quality. 

The two components of the system have been recognized 
separately in the Hipparcos raw data via a deviation from the distribution
of photon count rate as a function of time from the distribution expected
for a single star. For these component systems in the original Hipparcos
Catalog, the intermediate astrometric data are hard to interpret, since
the reference point for the data is not known a priori because it is a complicated
function of the photometric and geometric characteristics of the system,
and is different for the two data reduction consortia FAST and NDAC.

In the version of the Hipparcos Catalog by \cite{vanleeuwen97}, 
the system was recognized as a double star, but again solved  
with the standard five astrometric parameter model;
the astrometry given in the catalog seems to refer to component A, just
as in the original version of the Hipparcos Catalog. Our goal is to re-interpret the Hipparcos Intermediate Astrometric Data, 
taking the orbital solution derived in the present paper into account, 
and thereby revising the standard astrometric parameters and in 
particular the parallax. 

We used the Hipparcos Intermediate Astrometric Data (abscissae) from \cite{vanleeuwen07} 
and fitted corrections to the standard five astrometric parameters
(positions and proper motions in right ascension and declination, 
respectively, and the trigonometric parallax). Before the fitting, we
corrected the abscissae for orbital motion, with the orbital parameters given
in Table~3. Since the Hipparcos data are absolute astrometric measurements, 
the reference point for the orbital motion is the centre of mass of the system,
against which both components are seen to be moving. In order to derive the centre
of mass of the system, we used mass estimates of 0.9$\,$M$_\odot$ for the primary
and 0.7$\,$M$_\odot$ for the secondary. 

As expected, the largest adjustment occurs in the proper motions when the orbit
is taken into account in the analysis of the Hipparcos data. This is because
the Hipparcos measurements are spread out over approximately three years, whereas
the orbital period is more than a factor of 10 larger, so that the orbital phase
coverage of the Hipparcos measurements is relatively small. Thus, some orbital
motion has been presumably subsumed into the proper motions before, whereas the
corrected proper motions should be free of orbital motion. The new proper motions
are $\mu_{\alpha}*\cos{\delta} = -137.84 \pm 1.26$~mas\,yr$^{-1}$ and
$\mu_{\delta} = -33.53 \pm 1.45$~mas\,yr$^{-1}$; we reassess the kinematic membership 
of HD$\,$102077 to a moving group on the basis of these new proper motions in
Section~4.4. The parallax of HD$\,$102077 {of $20.59 \pm 2.14\,$mas}, however, does not change at all when the orbit is taken into
account in the fitting of the Hipparcos abscissae. 

We caution that we might have misinterpreted the reference point of the Hipparcos
Intermediate Astrometric Data, which would render our newly derived proper motions
meaningless. 
However, we have verified that the parallax result is very robust and does not
change at all even if we assume other reference points. So we conclude that the
published Hipparcos parallax is correct even in the case of orbital motion.

\subsection{Moving group membership}

The source HD$\,$102077 was proposed as candidate member of TW Hydrae \citep{Makarov01}. However, objections have been raised based on the space motion and weaker-than-expected lithium absorption \citep{Song02, Konig03}. We now revisit both points. 

Our new, high-resolution measurements of the Li EW confirm the lithium absorption value and the comparison of our data to stellar isochrones also suggests a higher age of HD$\,$102077 than is predicted for TW Hydrae.

\cite{Anders91} measured the radial velocity of HD$\,$102077 in July 1987 and \cite{Konig03} in January 2002; we measured the radial velocity in July 2013. The results are $15.9\,$km/s, $19 \pm 3\,$km/s, and $19.8 \pm {0.1} \,$km/s respectively. Another RV-measurement was done by {\cite{Nordstroem04}} who find $16.8 \pm {0.3}\,$km/s, but they do not give the epoch. These radial velocity values do not take the orbital motion into account. According to our orbit fit the relative motion of both components was $3.8\,$km/s in July 1987, $0.6\,$km/s in January 2002, and $5.0\,$km/s in July 2013. As in section 3.2.2 we adopt a maximum uncertainty of half the relative motion (e.g. $1.9\,$km/s for 1987). For the measurement by {\cite{Nordstroem04}} we assume an uncertainty of $2\,$km/s. The weighted mean of this values gives ${17.6 \pm 2} \,$km/s.

\begin{figure*}
\centering
\includegraphics[width=1\textwidth]{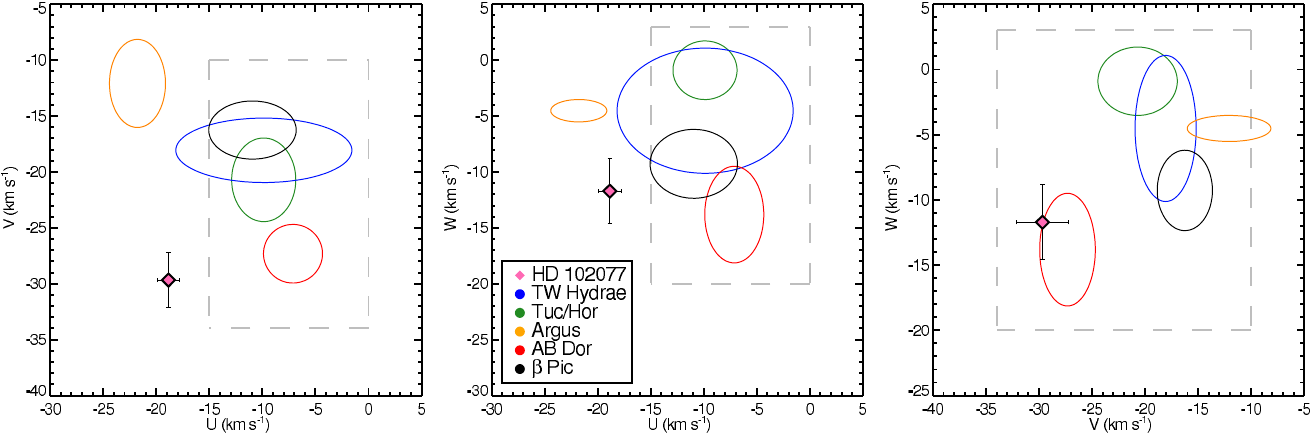}
\caption{2D projections of the UVW space velocities compared to well known moving groups,
including TWA. The dashed grey box is taken from \cite{Zuckerman04}
and shows the region of UVW space typically occupied by nearby young stars. The UVW velocities of the young moving groups are taken from \cite{Malo13}, the $2 \sigma$ uncertainties are plotted. HD$\,$102077 is clearly not a member of any of these groups.}
\label{movinggroup}
\end{figure*}

This RV value together with the newly derived proper motion can be used to calculate the space velocity as well as the kinematic membership likelihood. We derive UVWXYZ Galactic velocities and positions of the binary to be ${U=-18.9 \pm 1.1}\,$km/s, ${V=-29.7 \pm 2.5}\,$km/s, and ${W=-11.7 \pm 2.9}\,$km/s, as well as $X=17.7 \pm 1.4\,$pc, $Y=-44.0 \pm 2.4\,$pc, and $Z=10.1 \pm 2.3$pc and compare these values to those of TWA ($U=-9.9 \pm 4.2\,$km/s, $V=-18.1 \pm 1.4\,$km/s, ${W=-4.5 \pm 2.8}\,$km/s, and $X=12.5 \pm 7.1\,$pc, $Y=-42.3 \pm 7.3\,$pc, $Z=21.6 \pm 4.2$pc; \cite{Malo13}). We find that even though the Galactic position of HD$\,$102077 is close to most of the TW Hydrae members its space motion is different. The space motion of HD$\,$102077 {also does} not fit other young associations (see Fig.~\ref{movinggroup}). To calculate the moving group membership probability quantitatively we use BANYAN {II} (Bayesian Analysis for Nearby Young AssociatioNs {II}) \citep{Malo13, Gagne13} and find a probability of ${0.4\%}$ of HD$\,$102077 being a member of TWA. Even if the true parallax value happens to be outside of its $3 \, \sigma$ confidence region, the kinematic membership probability does not get higher than ${10 \%}$. Together with the estimated age this strongly suggests that HD$\,$102077 is not a member of TWA. The probability of HD$\,$102077 being a member of another young moving group is only non-zero for AB Doradus. With the newly derived system parameters it is about ${0.03 \%}$ and it is thus unlikely even though the age of the binary system matches the age of this moving group better. {HD$\,$102077 is most likely a member of the young Galactic field.}

\section{Summary}

Our photometric and spectroscopic measurements made it possible to study various aspects of the young, late type binary HD$\,$102077. We present a new, amended orbit fit with small uncertainties on the orbital elements. Using the revised Hipparcos parallax, we deduce a total system mass of $2.6 \pm 0.8\,$M$_{\sun}$ which leaves the possibility of HD$\,$102077 {having a third, close companion}. Further evidence for this hypothesis was found in the 2D cross correlation function of the two high-resolution spectra of HD$\,$102077. They show a large radial velocity difference that is in disagreement with the orbital fit, but which could be caused by a third component. {In addition, HD$\,$102077 probably has a wider companion at about $400\,$AU. We {also} use the derived orbit to revise the kinematics of the system. We find a proper motion of $\mu_{\alpha}*\cos{\delta} = -137.84 \pm 1.26$~mas\,yr$^{-1}$ and $\mu_{\delta} = -33.53 \pm 1.45$~mas\,yr$^{-1}$, as well as a radial velocity of ${17.6 \pm 2}\,$km/s. 

From the i'-z' colours we determine the spectral type of both components to be $K0 \pm 1$ and $K5 \pm 1$, respectively. We then use this result together with the individual Tycho B$_T$ and V$_T$ colours to estimate an age of the system. We find that HD$\,$102077 is very likely older than $25\,$Myr, which is in agreement with the lithium equivalent width of $200 \pm 4\,$m\AA $\,$  and with the absence of circumstellar material. {Our results are summarized in Tables~\ref{orbit} and~\ref{stellar}.}

With the amended kinematic data and the age estimate, we deduce the probability of HD$\,$102077 being a member of one of the young moving groups and find no match. Even the proposed candidate membership of HD$\,$102077 to the TW Hydrae Association is very unlikely.

For future investigation it would be interesting to study the radial velocity orbit of HD$\,$102077 to clearly answer the question whether the system contains a third component. 

\begin{acknowledgements}

We thank the staff at La Silla observatory for their support as well as Carolina Bergfors for observing 3 epochs of HD$\,$102077 and Felix Hormuth for his help with the analysis of the Lucky Imaging data. In addition, we thank Chad Bender for sharing his 2D cross-correlation code.

\end{acknowledgements}

\end{document}